\documentclass[twocolumn,aps,prl,showpacs,superscriptaddress]{revtex4-1}
\usepackage[latin9]{inputenc}
\usepackage{amsmath}
\usepackage{amssymb}
\usepackage{graphicx}
\usepackage{ulem}

\makeatletter
\usepackage{color}
\usepackage{bm}
\usepackage{float}

\makeatother

\begin{document}

\title{Monoenergetic High-energy Ion Source via Femtosecond Laser\\ Incident
Parallel to a Microplate}

\author{X. F. Shen}

\affiliation{Institut f\"ur Theoretische Physik I, Heinrich-Heine-Universit\"at D\"usseldorf,
40225 D\"usseldorf, Germany}

\author{A. Pukhov}
\email[Correspondence should be addressed to: ]{pukhov@tp1.uni-duesseldorf.de}

%\selectlanguage{english}%
%\selectlanguage{english}%

\affiliation{Institut f\"ur Theoretische Physik I, Heinrich-Heine-Universit\"at D\"usseldorf,
40225 D\"usseldorf, Germany}

\author{B. Qiao}

\affiliation{Center for Applied Physics and Technology, HEDPS, SKLNP, and School
of Physics, Peking University, Beijing, 100871, China}

\date{\today}

\begin{abstract}
Using fully three-dimensional particle-in-cell simulations, we show
that readily available femtosecond laser systems can stably generate
proton beams with hundred MeV energy and low spread at $\sim1\%$
level by parallel irradiation of a tens of micrometers long plasma
plate. As the laser pulse sweeps along the plate, it drags out a huge
charge ($\sim$100 nC) of collimated energetic electrons and accelerates
them along the plate surface to superponderomotive energies. When
this dense electron current arrives at the rear end of the plate,
it induces a strong electrostatic field. Due to the excessive space
charge of electrons, the longitudinal field becomes bunching while
the transverse field is focusing. Together, this leads to a highly
monoenergetic energy spectrum and much higher
proton energy as compared to simulation results from typical target normal
sheath acceleration and radiation pressure acceleration at the same laser parameters. 
\end{abstract}


\maketitle

The research of laser-driven ion acceleration received renewed interest
in recent years due to several breakthroughs achieved in experiments
\cite{Gaillard2011,Wagner2016,Kim2016,Higginson2018} and the forthcoming petawatt laser
devices \cite{Danson2015}. The maximum proton energy has been improved
from 58MeV \cite{Snavely2000} to 94MeV \cite{Higginson2018} with
advancements in both laser technology and targetry. However, the ion
beams still exhibit an exponentially decaying energy distribution.
This is a major drawback to cancer therapy and other applications,
which require energy spread only about 1$\%$ \cite{Bulanov2002,Macchi2013,Daido2012}.
An energy-selection system out of a broad energy spectrum ion source sophisticates the device and leads to huge particle loss \cite{Linz2016,Masood2014}.
Meanwhile, most of experiments are accomplished on large laser
facilities which deliver 100s J energy within a picosecond at low
repetition rate \cite{Gaillard2011,Wagner2016,Higginson2018,Snavely2000}. These lasers can be operated only in a few national
laboratories \cite{Danson2015}. Compared to this, a high-repetition-rate,
low-cost and stable femtosecond laser is more preferable for developing
the future compact ion sources. Actually, 100 Terawatt (TW)-class
femtosecond laser systems have been distributed widely around the
world and also many multi-petawatt (PW) ones are currently operational,
under construction or in the planning phase \cite{Danson2015}, for
which generation of monoenergetic high-energy ion beams is one of
primary applications. Nevertheless, in the present femtosecond laser-ion
acceleration experiments, proton energies are typically much lower than those obtained from picosecond laser pulses
and the energy spectra are also broad \cite{Henig2009,Dollar2012,Schramm2017,Scullion2017,Bin2018,Dover2020}.

To achieve monoenergetic ion beams, a longitudinal bunching accelerating
field is important, in which fast ions experience a smaller field
and the slow ones a larger field. In traditional radio-frequency accelerators,
such a bunching field is realized through controlling the phase of
the synchronous particle relative to the crest of the accelerating
wave \cite{Wangler1998}, while in laser-ion acceleration, it appears
as a longitudinal negative gradient electric field acting on the accelerated
ions. In one of the mostly investigated laser-ion acceleration mechanisms,
radiation pressure acceleration (RPA) \cite{Esirkepov2004,Robinson2008,Qiao2009,Shen2017},
such a bunching field was supposed to exist through piling excessive
electrons at the rear surface. However, to obtain monoenergetic ion beams, the
RPA requires ultraintense laser pulses ($>10^{22}\,{\rm W/cm^{2}}$)
and a large spot size simultaneously, which remains a big challenge
in experiments even for multi-PW lasers. In addition,
the RPA is plagued by strong electron heating due to effects of transverse
instabilities \cite{Pegoraro2007} and finite spot size \cite{Dollar2012}.
These may induce relativistic transparency and destroy the bunching
electric field. As a consequence, the obtained ion energy is
rather limited and energy spread is very large.

In the other widely studied mechanism, target normal sheath acceleration
(TNSA) \cite{Pukhov2001,Wilks2001,Schwoerer2006,Nakatsutsumi2018},
this longitudinal bunching field is absent due to the low density
of energetic electrons. The energy spectrum of TNSA ions is characterized
by an exponential decay. Moreover, 
the acceleration time is related to laser pulse duration, which undoubtedly
leads to much lower energy with femtosecond laser pulses \cite{Mora2003,Fuchs2006}.

Many efforts have been devoted to overcoming the limitations of these
mechanisms to improve the ion beam parameters, especially the
maximum energy and energy spread. Here, we may mention
target designs \cite{Bin2018,Schwoerer2006,Chen2009,Bartal2012},
multi-pulse schemes \cite{Toncian2006,Markey2010}, post-acceleration
\cite{Kar2016} and novel mechanisms \cite{Hilz2018,Mackenroth2016,Brantov2016,Matsui2019,Shen2019,Ziegler2020}.
Nevertheless, the obtained energy spreads are still larger than 10$\%$
due to the absence of a self-established bunching field, and also
the increasing complexity in the laser pulse and target configurations
may reduce the repetition-rate and robustness of experiments. Therefore,
the future of producing high-energy (like 100MeV protons), monoenergetic
(energy spread about 1$\%$) ion beams with the known mechanisms is
vague.

\begin{figure}
\includegraphics[width=8.6cm]{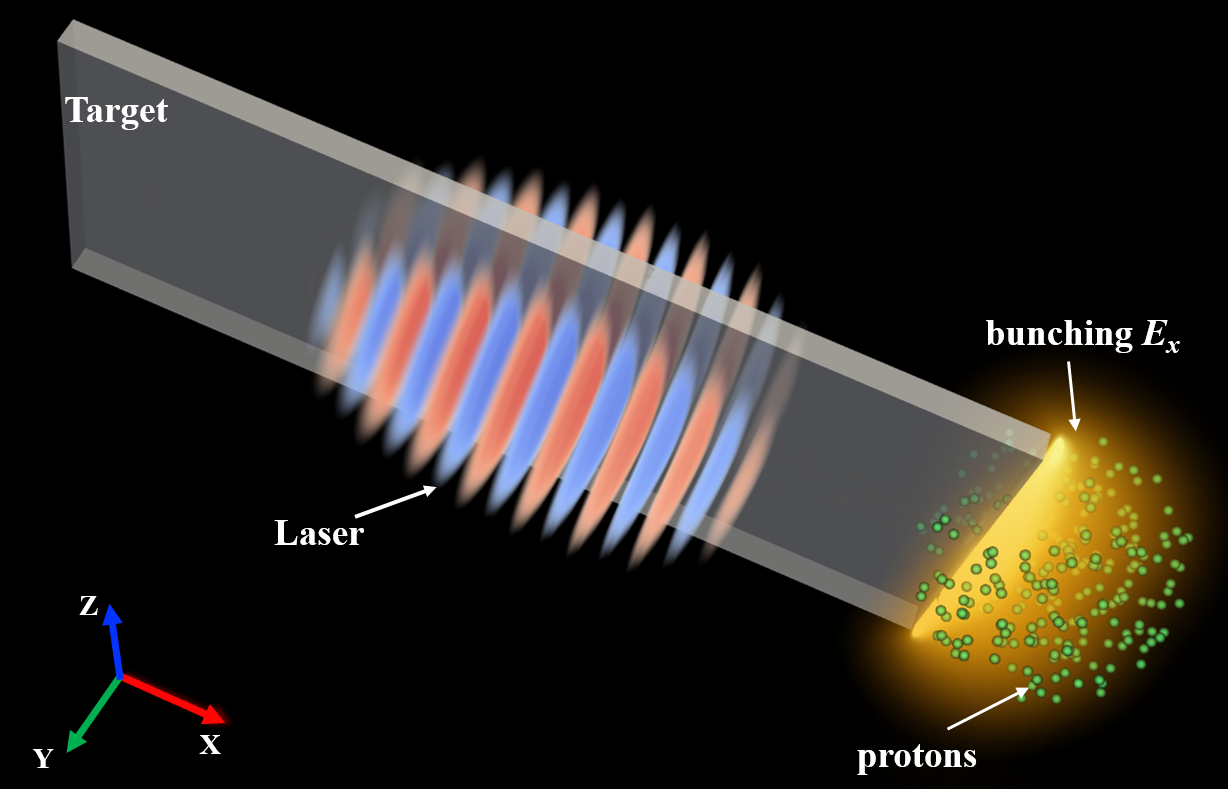} \caption{(color online) Schematic of ion acceleration mechanism via a femtosecond
laser (red-blue) incident parallel to a micro-scale plate (grey).
The yellow domain shows the distribution of the longitudinal bunching
field and green dots represent accelerated ions. Here the schematic does not represent the true scale, where an infinitely long plate along $z$ direction can be used in experiment.
}
\label{fig:schematic} 
\end{figure}

In this Letter, we show that a longitudinal bunching accelerating
field is built spontaneously when a currently available femtosecond
laser pulse is incident on an edge of a simple micro-scale plasma
plate, parallel to its surface, as shown by Fig. \ref{fig:schematic}.
As the laser pulse of intensity $>10^{20}$ ${\rm W/cm^{2}}$ sweeps
along the plate, it extracts abundant buckets of electrons from
the plate into vacuum and accelerates them forward along the plate
surface to superponderomotive energies via direct laser acceleration
(DLA) \cite{Pukhov1999} and surface plasma wave (SPW) \cite{Riconda2015,Pitarke2007}.
The charge of energetic electrons (many tens to hundreds of nC) injected
into the accelerating region at the rear edge of the plate is much
larger than that of protons placed there (just a few nC). This ensures
that the protons near the laser propagation axis are surrounded by
the excessive space charge of electrons, that leads to a negative
gradient longitudinal bunching field and also a transverse focusing
field. Finally, a quasi-monoenergetic proton beam with peak energy $>$100MeV,
energy spread about 1$\%$ and particle number $\sim10^{9}$ can be
stably obtained.

\begin{figure}[b]
\includegraphics[width=8.6cm]{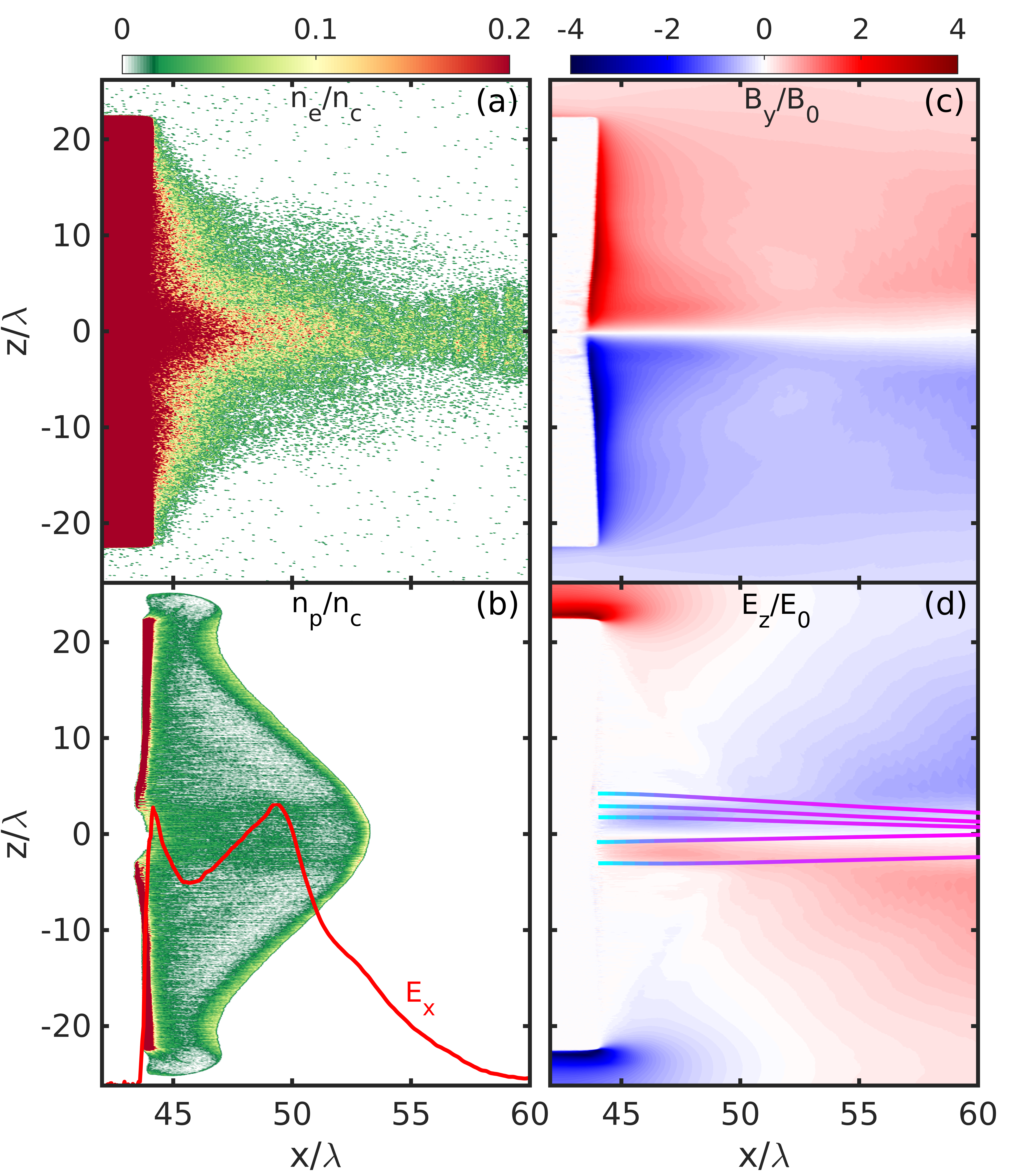} \caption{(color online) (a) and (b) show the distributions of electron and
proton density in ($x,z$) plane at $t=82T_{0}$, respectively. The
red line in (b) displays the on-axis profile of the accelerating field
$E_{x}$. (c) and (d) show the transverse magnetic field $B_{y}$
and electric field $E_{z}$, respectively. The lines in (d) correspond
to the collimated trajectories of selected protons, where the color
represents the proton energy.}
\label{fig:fig2} 
\end{figure}

The three-dimensional (3D) particle-in-cell (PIC) simulations are
conducted with the EPOCH code \cite{Arber2015} and the VLPL code
\cite{VLPL_CERN}. The simulation box is 120$\lambda\times40\lambda\times52.5\lambda$
in the $x\times y\times z$ directions, containing $2400\times1600\times1050$
cells, respectively. Here, a higher resolution in $y$ direction
is used to resolve the plasma skin depth and the process of electrons
extraction out from the plate, since a $y$-polarized laser pulse
is chosen. We use larger box length in $z$ direction because of a
large plate to mimic the real experimental situation. The laser intensity
is $I_{0}=7.8\times10^{20}{\rm W/cm^{2}}$ with the wavelength $\lambda=800$nm.
The laser pulse has the Gaussian profile in both space and time with
the radius $r_{L}=7.5\lambda$ and pulse duration $\tau_{L}=45$fs,
respectively, which is focused at the front edge of the plate. A plasma plate of high-Z material (we assumed gold)
had the dimensions $x\times y\times z=43.75\lambda\times0.75\lambda\times45\lambda$.
Its rear end has been covered by a hydrocarbon (CH) layer. To reduce
the computational resources, the electron densities for the plasma plate and CH layer are both
chosen as relativistically overdense $n_{e}=30n_{c}$. Here $n_{c}=\pi m_{e}c^{2}/e^{2}\lambda^{2}$ is the
critical density.  
The initial charge states of ions are given according
to the Ammosov-Delone-Krainov formula \cite{Ammosov1986}, which means
the ion species are Au$^{51+}$ in the main target, C$^{6+}$ and
H$^{+}$ in the hydrocarbon layer.
The transverse dimensions of the CH layer are $0.75\lambda\times45\lambda$ (same as the plate), while the longitudinal is set to $l_x=0.4\lambda$ to ensure the proton charge realistic. 
The density ratio of proton to carbon ion is $n_{p}:n_{C^{6+}}=1:1$. 
The
macro-particles in each cell for electrons, gold ions, carbon ions
and protons are 8, 1, 8 and 32, respectively. Open boundary conditions
for fields and particles are employed. The numerical convergence has
been confirmed via comparing the interested physical quantities with
simulations at different resolutions and the two codes. 

Figure \ref{fig:fig2}(a) shows the electron density distribution
in ($x,z$) plane at t=$82T_{0}$, where $t=0$ represents the time
when the pulse peak enters the simulation box. One sees that the electron
beam exiting the plate is highly collimated. This is because, as the
laser pulse sweeps along the plasma plate, electrons within the skin
depth are continuously ripped off by the laser electric field $E_{y}$
and accelerated forward via the mixed DLA and SPW mechanisms. Both
of them contribute to the increments of $p_{x}$, resulting in $p_{x}\gg p_{\perp}\gg m_{e}c$
\cite{Pukhov1999,Pukhov2002,Riconda2015}. For DLA, it is due to the
${\bf v}\times{\bf B}$ force and for SPW, it is the longitudinal
surface plasmon field \cite{Kluge2012}. The SPW can be easily excited when a laser pulse is incident along an overdense plasma surface with a sharp edge, and then travels along the plate with velocity close to $c$ \cite{Riconda2015,Pitarke2007}. Such an acceleration benefits from a long plate irradiated by the center of laser pulse. It promises longer acceleration distance and larger electron charge compared to a grating target, and it does not require specific modulations \cite{Fedeli2016,Chopineau2019}. After electrons are injected into
the vacuum, a transverse focusing magnetic field $B_{y}$ is induced,
as shown by Fig. \ref{fig:fig2}(c), which offsets the defocusing
force of Coulomb field $E_{z}$ {[}shown by Fig. \ref{fig:fig2}(d){]}.

\begin{figure}[b]
\includegraphics[width=7.5cm]{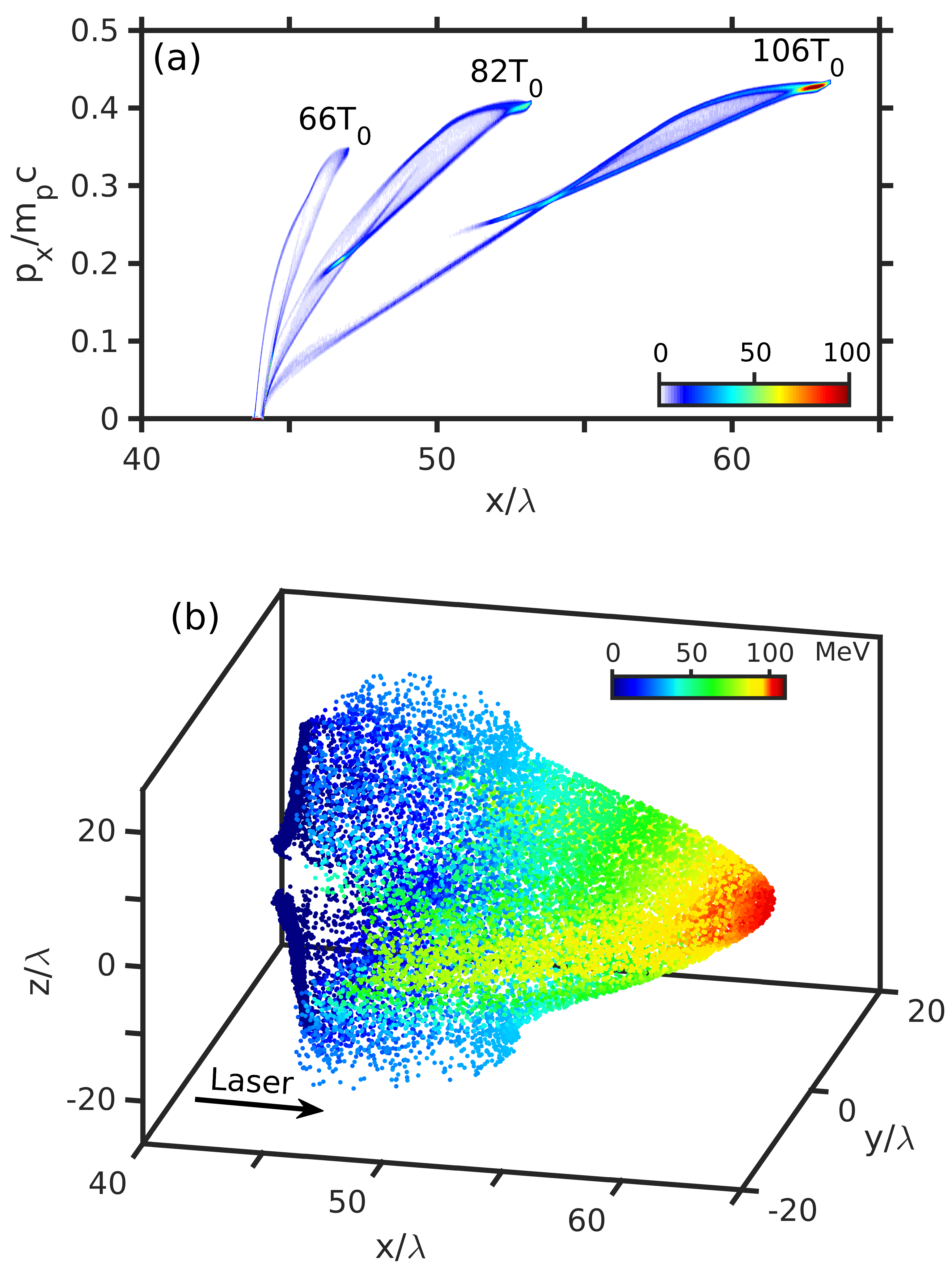} \caption{(color online) (a) Evolution of proton phase space, where the protons
are selected within a 10$^{\circ}$ divergence angle and the color
represents the relative proton number. (b) 3D image of the proton
distribution at $t=106T_{0}$, where the color shows the energy and
the local density of dots indicates that of proton. Each dot represents
200 macro-particles. 
}
\label{fig:fig3} 
\end{figure}

\begin{figure}[t]
\includegraphics[width=8.6cm]{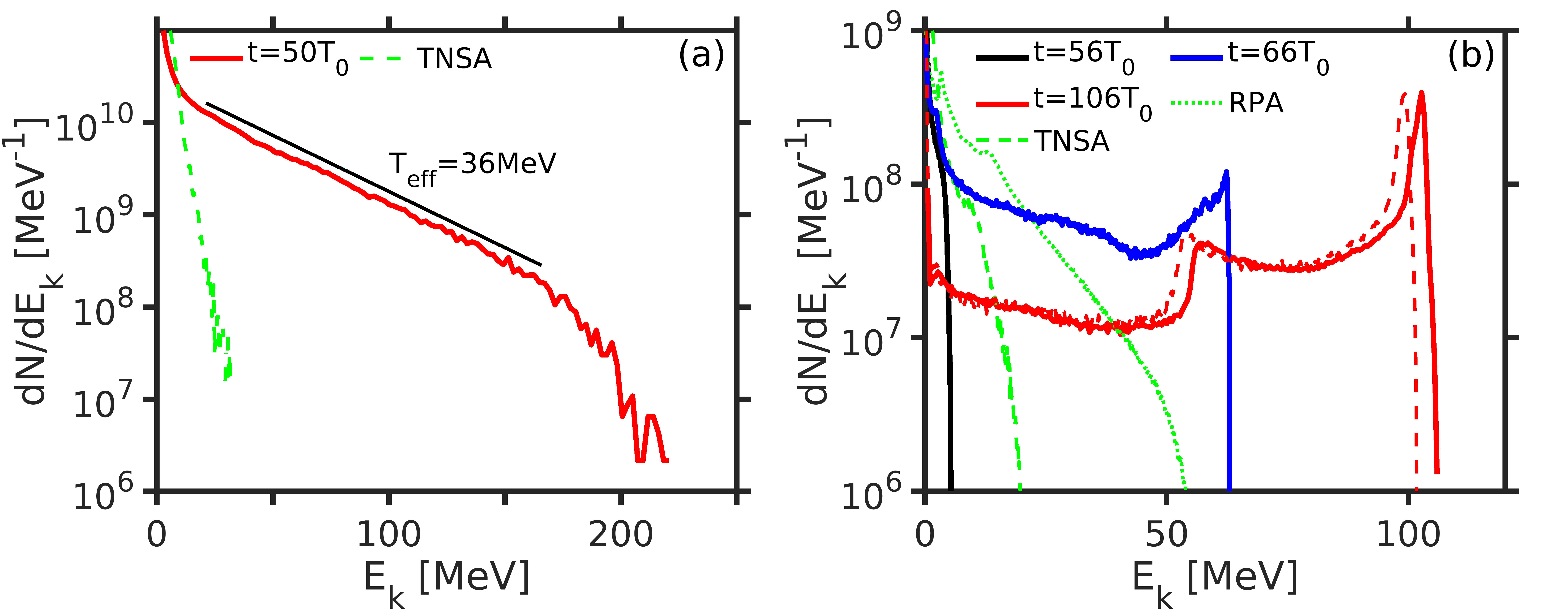} \caption{(color online) Energy spectra of electrons (a) and protons (b). In
(b), the black, blue and red solid lines show the evolution
of proton energy spectra in our scheme at $t=56T_{0}$, $66T_{0}$ and $106T_{0}$, respectively. The red dashed line displays the results with a misalignment of $2\lambda$ along $y$ direction. Moreover, the green dashed
lines in (a) and (b) represents the corresponding energy spectra obtained
from typical TNSA, while the green dotted line in (b) shows the results from RPA, 
where the electron and proton number multiply
a factor of 0.1 and 0.01, respectively, to make it suitable to the
coordinate range.  Note that here protons are selected inside a $10^{\circ}$
divergence angle.}
\label{fig:fig4} 
\end{figure}

The space charge of this collimated electron beam is huge. When the
laser pulse reaches the rear surface, the charge of high energy electrons ($\gamma_{e}>10$)
outside of the plate is about 150nC ($\sim10^{12}$), which accords
well with the estimation of $N_{e}=4r_{L}l_{s}l_{p}n_{e}$, where
$l_{p}$ is the plate length, $l_{s}=c/\omega_{pe}$ is the skin depth
and $\omega_{pe}=\sqrt{4\pi n_{e}e^{2}/m_{e}}$ is the plasma frequency.
Then, a large part of them are continuously injected into the ion
acceleration region, where the electron space charge can reach about
50nC, much larger than the charge of protons (about 2.7nC within the
focus spot). This ensures that protons, especially those around the
propagation axis, are surrounded by a negative electron cloud during
the acceleration process. The distribution of proton density in ($x,z$)
plane is shown in Fig. \ref{fig:fig2}(b), which is much lower than
the electron density {[}Fig. \ref{fig:fig2}(a){]}, even at the compressed
density peak. In one dimensional situation, $\partial E_{x}/\partial x=-4\pi e(n_{e}-n_{p})$.
Therefore, a longitudinal bunching electric field forms around protons, 
as the red line shown in Fig. \ref{fig:fig2}(b),
like the ``compressed electron layer\char`\"{} in RPA \cite{Yan2008,Macchi2009},
which compresses the proton phase space {[}Fig. \ref{fig:fig3}(a){]}
and reduces the energy spread {[}solid lines in Fig. \ref{fig:fig4}(b){]}
repeatedly. This is the reason for the formation of a proton density
peak in Fig. \ref{fig:fig2}(b).

Meanwhile, for protons, the $E_{z}$ field provides a focusing force,
while $B_{y}$ is defocusing. However, since $E_{z}$ is comparable
to $B_{y}$ and the velocities of protons $v_{x}$ are
smaller than the light speed, protons are actually focused by these
transverse fields. We show trajectories of some selected protons in
Fig. \ref{fig:fig2}(d), where the color marks the evolution of proton
energy. Moreover, we can also see more protons gather around the $x$-axis in Fig. \ref{fig:fig2}(b). Such a self-established longitudinal
bunching and transverse focusing field configuration makes our scheme
robust and suitable for generation of monoenergetic ion beams. This
is the key difference from the typical TNSA, where a debunching field
dominates the acceleration process. Furthermore, compared
to the requirement of a fragile balance condition in RPA \cite{Shen2017,Pegoraro2007},
the bunching field in our scheme is self-established
on the basis of large number of high-energy electrons, which is very
robust only if the plate is long enough to provide sufficient charge
of electrons.

Figure \ref{fig:fig3}(b) shows the 3D perspective view of proton
distribution at the end of acceleration, where the color marks the
proton energy and the local density of dots indicates that of proton,
since each dot was randomly selected and has the same weight. We see
that the highest energy protons gather around the propagation axis
(red region), forming a high-density ($\sim0.1n_{c}$), high-energy
($>100$MeV) quasi-monoenergetic proton bunch.

Further, the red line in Fig. \ref{fig:fig4}(a) represents the electron
effective temperature $T_{\rm eff}$ at $t=50T_{0}$, which is about 36MeV, much larger
than the value of 6.4MeV given by ponderomotive scaling $T_{\rm pond}=(\sqrt{1+a_{0}^{2}/2}-1)m_{e}c^{2}$
\cite{Pukhov1999,Wilks1992}. The electron density is also 
high, about $n_{h}=2.5n_{c}$, when electrons are attracted to flow
into the mid-plane from both sides. The initial longitudinal
field $E_{sh}=\sqrt{8\pi n_{h}T_{\rm eff}/e_{{\rm N}}}=4.54\times10^{13}\,{\rm V/m}$
with $e_{{\rm N}}\approx2.71828$ \cite{Mora2003}, is consistent with the simulation
result $5.0\times10^{13}\,{\rm V/m}$, which is almost comparable
to the peak value of laser field. Due to the collimation and large
longitudinal recirculation radii of high-energy electron beams, $E_{x}$
decays only slowly and the acceleration of the high-energy protons
could last for 150fs (about four times longer than $\tau_L$). This strong, long-lasting accelerating
field leads to a high efficient acceleration, which also explains
why almost all the protons near $x$-axis are evacuated in Fig. \ref{fig:fig2}(b).

Figure \ref{fig:fig4}(b) shows the energy spectra of proton beam
at different times. At the beginning, the energy spread is large (black
line), since protons feel a positive gradient $E_{x}$ as they are
pulled out from the CH layer. Subsequently, the energy spread decreases
constantly due to the longitudinal bunching field caused by excessive
electrons, as shown by the blue and red lines. Finally, a
high-energy quasi-monoenergetic proton beam, with peak energy $>$100MeV,
energy spread about $1.17\%$ and particle number $8\times10^{8}$
(0.13nC) within the peak (FWHM), is obtained (red line). This 
low energy spread persists for a long time since the further contribution
of Coulomb explosion could be ignored considering the co-propagating,
collimated electron beam. Note that our results is very robust, which is not sensitive to a possibly slight misalignment [red dashed line in \ref{fig:fig4}(b)], variation of target dimensions (length, thickness or height) \cite{Supplementary}, and existence of preplasma \cite{Supplementary}. 

Moreover, as a baseline comparison, we also
performed 3D simulations with a laser pulse obliquely incident on
a flat target with incidence angle 30$^{\circ}$,  
where the laser parameters keep the same. The electron
and proton energy spectra are shown by the dashed green lines in Fig.
\ref{fig:fig4}(a) and \ref{fig:fig4}(b), respectively. The effective
temperature is much lower and the maximum proton energy is about five
times less than that in our scheme and the energy spectrum is broad. Furthermore, the green dashed-dotted line shows the results from a target with area density satisfying the optimal condition of RPA \cite{Macchi2009}. Due to the effects of instabilities \cite{Pegoraro2007}, the energy spectrum is exponentially decaying and the maximum energy is only half of ours.

\begin{figure}[b]
\includegraphics[width=7cm]{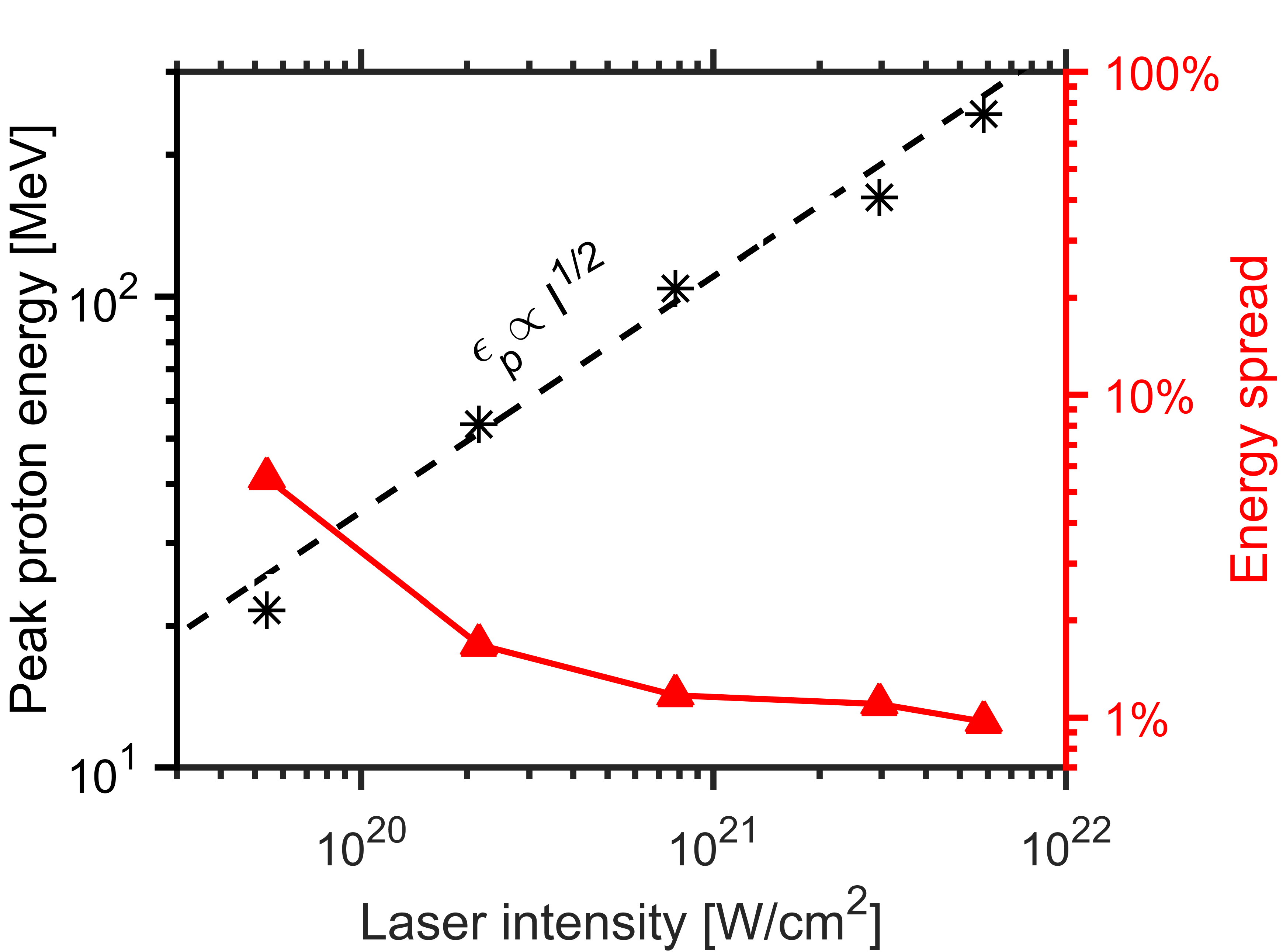} \caption{(color online) The peak proton energy $\epsilon_{p}$ (black asterisks)
and energy spread (red triangles) with varying laser intensity, where
the other parameters are kept almost the same as those of Fig. 2,
except the electron density to avoid the relativistic transparency.
The dashed black line displays the best fit scaling for peak proton
energy.}
\label{fig:fig5} 
\end{figure}

Figure \ref{fig:fig5} illustrates the peak proton energy $\epsilon_{p}$
(black asterisks) and energy spread (red triangles) as a function
of laser intensity $I_{0}$ obtained from 3D simulations, where the
other parameters keep almost the same. The numerical results suggest
that the scaling of $\epsilon_{p}$ satisfies $\epsilon_{p}\sim\alpha(I_{0}/I_{18})^{1/2}$,
with coefficient $\alpha\approx3.5\,{\rm MeV}$ and $I_{18}=10^{18}\,{\rm W/cm^{2}}$.
Though the scaling is similar to that of TNSA, the proportionality
factor $\alpha$ is significantly higher, as we discussed before.
This stems from the high-quality electron beam characterized with large particle number, high effective temperature
and small divergence angle. More importantly, as the red triangles
shown in Fig. \ref{fig:fig5}, the energy spread always stays at an
extremely low level, and even with moderate intensity 
$I_{0}<10^{20}\,{\rm W/cm^{2}}$, it is still less than $10\%$. Proton
beams with an energy spread at $1\%$ level can be stably obtained
as the laser intensity increases. To the best of our knowledge, no
experiments or 3D PIC simulations of laser-ion acceleration have reported
such high-quality ion beams \cite{Macchi2013,Daido2012,Qiao2019}.
Though in Ref. \cite{Haberberger2012}, proton beams with energy spread
1$\%$ were achieved, but only $\sim10^{5}$ protons within
peak and the experiments are performed with CO$_{2}$ laser systems.

In conclusion, a robust scheme for achieving 100MeV proton beams with
energy spread at the $\sim1\%$ level is proposed, where
a longitudinal bunching and transverse focusing field is self-established
through irradiating a femtosecond laser pulse parallel to a micro-scale
plate. This novel interaction geometry not only promises a high-quality proton beam, but also enables other fundamental studies and various applications. For example, as the high-energy electrons move forward, a large return current and therefore a strong quasistatic magnetic field is induced, which could compress the plasma to an ultrahigh density \cite{Supplementary,Kaymak2016}. Therefore, a microscale ultradense $Z$-pinch can be expected to form.

\section*{Acknowledgements}

This work is supported by the DFG (project PU 213/9) and GCS J\"ulich
(project QED20). X.F.S. gratefully acknowledges support by the Alexander
von Humboldt Foundation. X.F.S. acknowledges helpful discussions with
Ke Jiang at HHU.\bigskip{}

\end{document}